\documentclass[letterpaper,10pt]{article}
\usepackage[runin]{abstract}
\usepackage{graphicx}
\usepackage{algorithmic}
\usepackage{algorithm}
\usepackage{url}
\usepackage{amsmath,amssymb,amsthm}
\usepackage{multirow}
\usepackage{hyperref}

\newcommand \mb[1] {\boldsymbol{#1}}
\def\R {\mathbb{R}} %
\def\d {\mathrm{d}}
\def\P {\mathbb{P}}

\hypersetup{%
	pdftitle={GPU in Physics Computation: Case Geant4 Navigation},%
	pdfauthor={},%
	pdfsubject={},%
	pdfkeywords={},%
	pdfborder={0 0 0},%
	colorlinks={false}%
}

\begin{document}

\begin{center}
\section*{GPU in Physics Computation: \\ Case Geant4 Navigation}

\vspace*{1em}
\textsc{Otto Seiskari, Jukka Kommeri, Tapio Niemi}%

\emph{otto.seiskari@aalto.fi, \{kommeri, tapio.niemi\}@cern.ch}\\~\\
Helsinki Institute of Physics, Technology Programme,\\ 
CERN, CH-1211 Geneva 23, Switzerland
\\[1em]

January 16, 2011
\\[2em]

\end{center}

\begin{abstract}
General purpose computing on graphic processing units (GPU) is a
potential method of speeding up scientific computation with low cost
and high energy efficiency. We experimented with the particle physics
simulation toolkit Geant4 used at CERN to benchmark its geometry
navigation functionality on a GPU. The goal was to find out whether
Geant4 physics simulations could benefit from GPU acceleration and how
difficult it is to modify Geant4 code to run in a GPU.

We ported selected parts of Geant4 code to C99 \& CUDA and
implemented a simple $\gamma$-physics simulation utilizing this code
to measure efficiency. The performance of the program was tested by
running it on two different platforms: NVIDIA GeForce 470 GTX GPU
and a 12-core AMD CPU system. Our conclusion was that GPUs can be a
competitive alternate for multi-core computers but porting existing
software in an efficient way is challenging.
\end{abstract}

\section{Introduction}

Graphic processing units (GPUs), also known as graphic adapters, or 3D
cards, were originally expansion cards for personal computers. The
purpose of these devices is to externalize graphics related computing
from the main processor (CPU) and to increase the speed of these
calculations with specialized hardware. 
The main driving force of GPU technology development has been the
demand for fast 3D graphics in computer games. Polygon-based rendering
is an ideal application for specialized parallel co-processor
architectures. Subsequently, more complex and customizable computation
capabilities have been added, and recently, GPUs have developed into
fully programmable general purpose parallel
computers \cite{luebke2006gpgpu}.

The modern GPUs can achieve high computation speeds, measured in
floating point operations per second (FLOPS), compared to their size,
cost and energy consumption. This makes them attractive for use also
outside the computer game industry. These devices, and similar
GPU-derived accelerators with no graphic processing capabilities, have
recently developed interest in many fields of scientific computing and
have been successfully used to speed up various
applications.~\cite{jahnke2008gpu}

The high computing performance on GPUs is based on massive parallelism
and is generally achievable only if the computation can be split into
a large number of parallel tasks that require little global
communication.
GPU programming can be done using relatively high-level languages and
toolkits with varying level of portability. Alternatives include
NVIDIA's CUDA \cite{cuda} and the newer cross-platform
OpenCL \cite{opencl} by Khronos group.

In the current work we study
whether GPUs are beneficial for particle physics simulations and how
existing physics software can be ported to GPUs. The motivation for
our research is the Large Hadron Collider particle accelerator that
started running at CERN during in late 2009 and it is now producing
data for physic experiments. The purpose of this accelerator is to
provide new information on the basic structure of matter. From
computing and data storage point of view, LHC experiments produce over
15 petabytes of data annually.

\section{Background}

\subsection{Monte Carlo Simulations in Particle Physics}

Particle physics simulations have an essential role in contemporary
experimental nuclear and particle physics \cite{geant4}. They are used
in designing and calibrating particle detectors and generating
detectable signatures for hypothesized physics phenomena. Modern
particle physics experiments require increasingly large amounts of
Monte Carlo data from complex and computationally intensive
simulations \cite{geant4}. Examples of physics experiments requiring
extensive Monte Carlo simulations include CMS \cite{cms} and ATLAS
\cite{atlas} experiments related to the Large Hadron Collider (LHC)
\cite{lhc} at CERN, Geneva, Switzerland.

Monte Carlo particle physics simulations are also important in other fields, such as
medical physics \cite{zaidi1999relevance,geant4,jahnke2008gpu}, where
an example application is the planning of maximally safe and accurate
radiotherapy treatments.

\subsection{Geant4}

Geant4 \cite{geant4} is ``a toolkit for simulating the passage of particles through
matter''. It can simulate a comprehensive set of different physical
processes over a wide range of particle energies for a variety of
long-lived particles. It is designed to offer a comprehensive set of
functionalities needed to construct a (Monte Carlo) particle physics simulation
program for a specific setup. It is a large object oriented system
implemented in C++.

The Geant4 toolkit is used in many contemporary particle physics
experiments, such as CMS \cite{cms} and ATLAS \cite{atlas} at CERN and
BaBar \cite{babar} at SLAC. It is also utilized in medical
applications, for example in radio therapy
planning \cite{jahnke2008gpu}.

\subsection{The NVIDIA Fermi GPU Architecture}

This is a brief description of the hardware architecture (NVIDIA ``Fermi'') of the GPU used in our experiments.
Its differences compared to contemporary multi-core CPU architectures are a key factor determining the performance of GPU programs.

In the Fermi architecture \cite{cuda}, a \emph{device} (e.g. one
GPU) consists of several (e.g. 16 on the Tesla GPUs) \emph{streaming
  multiprocessors} and each multiprocessor has a number (32) of
\emph{cores}. Each core executes a \emph{warp} of threads (32 threads)
in a manner called \emph{Single-Instruction, Multiple-Thread} (SIMT)
by NVIDIA. In SIMT, the threads in one warp always execute a common
instruction with different data. If the instruction paths of the
threads diverge on a data-dependent branch instruction, the divergent
branches are executed sequentially, that is, threads in one branch
wait for the threads in the other branch to execute. SIMT can be
viewed as a special case of the SIMD (Single-Instruction, Multiple-Data)
architecture.

The device has a large \emph{global memory} with unsynchronized
read-write access from the multiprocessors. There is also cached
read-only memory, \emph{constant memory} and special \emph{texture
  memory} with automatic interpolation. Each multiprocessor has a
smaller amount of \emph{shared memory} that can be used manually for
local synchronization or automatically as L1 cache for the global
memory. In addition, each multiprocessor has a set of 32-bit
\emph{registers}. The Fermi GPUs are capable of native 32-bit and
64-bit IEEE-compliant floating point calculations.

In order to efficiently exploit the full potential of a Fermi GPU, one
must be able to parallelize the computational problem to thousands of
threads that require little global synchronization. One
multiprocessor can execute $32\times32 = 1024$ threads concurrently
and to hide latencies from global memory access with context
switching, there should be multiple warps of suspended threads waiting
on each multiprocessor. There are also other constraints, such as
the limited number of registers on each multiprocessor.

Because of these constraints and the nature of SIMT architecture, algorithms that can be implemented efficiently on parallel MIMD (Multiple-Instruction, Multiple-Data) machines, such as multi-core CPUs,
do not necessarily execute efficiently on NVIDIA GPUs.

\section{Methods}

We study a minimalistic Monte Carlo physics simulation program for calculating where $\gamma$-particles interact with matter. A similar program could, in theory, be used as an GPU accelerated part of a more complex physics framework. The most computationally intensive parts of this program are the geometry \emph{navigation} calculations (see section \ref{section-geant4-navigation}), that are done using Geant4 code.

\subsection{The Simple Particle Physics Simulation}

\begin{algorithm}
 \caption{Monte Carlo simulation for free path lengths of particles\label{algorithm-monte-carlo}}
 \begin{enumerate}
 \item Generate $W \sim \mbox{Uniform}((0,1])$
 \item Set $Y = -\log W$ (now $Y$:s have distribution $\mbox{Exp}(1)$)~\cite{geant4}.
 \item Calculate $T = G_{\mb x_0,\mb \omega}^{-1}(Y)$ as follows:
\begin{itemize}
\item[] Set $S_0 = 0,\ t_0 = 0,\ j \leftarrow 0$
\item[] While $S_j < Y$, do
\begin{itemize}
\item[] Set $j \leftarrow j+1$
\item[] Compute the next intersection $t_j > t_{j-1}$ and get $\mu_j$ using Geant4.
\item[] Set $S_j = S_{j-1} + (t_j - t_{j-1})\mu_j$ (may be $\infty$)
\end{itemize}
\item[] Set $T = t_{j-1} + \frac{Y - S_{j-1}}{\mu_j}$ 
\end{itemize}
\end{enumerate}
\end{algorithm}

Suppose a beam of $N(0)$ particles moves through matter. The number of particles left after distance $x$ can be modeled with the differential equation
\begin{equation}
 \label{differential-equation-for-N}
 \frac{\d N(x)}{\d x} = -N(x) \mu(x),
\end{equation}
where $\mu = \sigma \rho N_A / A$ is the mean free path in a material with density $\rho$, atomic weight $A$ and a total cross section of $\sigma$~\cite{green2000physics}.
The solution of \eqref{differential-equation-for-N} is
\begin{equation}
 N(x) = N(0) e^{-\int_0^x \mu(y)\d y} =: N(0) e^{-G(x)}
\end{equation}
and the probability of a single particle interacting before traveling distance $x$ is therefore
\begin{equation}
 \P(X < x) = 1 - \frac{N(x)}{N(0)} = 1 - e^{-G(x)},
\end{equation}
where $X$ is a random variable representing the free path length of the particle. Assuming $G$ is invertible, we can generate random variables $Z$ with this distribution from $\mbox{Exp}(1)$-distributed variables $Y$ as
\begin{equation}
 Z = G^{-1}(Y).
\end{equation} 

Suppose we have a geometry of objects defining an attenuation coefficient $\mu(\mb x)$ in a point $\mb x \in \R^3$, and a particle with starting point $\mb x_0$ and a unit direction $\mb \omega$. The probability of the particle interacting before distance $t \geq 0$ is
\begin{equation}
 \label{distribution-of-free-path-lengths}
 \P(T < t) = 1 - e^{-\int_0^t \mu(\mb x_0 + y \mb \omega) \d y} =: 1 - e^{-G_{\mb x_0,\mb \omega}(t)}.
\end{equation}
To generate random variables $T$ with the correct distribution, one must be able to compute $G_{\mb x_0,\mb \omega}^{-1}(t)$. Suppose the geometry is composed of a finite amount of different materials that define regions with piecewise smooth boundaries. Then we may calculate the integral as
\begin{equation}
 G_{\mb x_0, \mb \omega}(t) = \sum_{i=1}^M (t_i - t_{i-1}) \mu_i + (t - t_M) \mu_{M+1} =: S_M + (t - t_M) \mu_{M+1} 
\end{equation}
and
\begin{equation}
 G_{\mb x_0,\mb \omega}^{-1}(z) = t_M + \frac{z - \sum_{i=1}^M (t_i - t_{i-1}) \mu_i}{\mu_{M+1}} = t_M + \frac{z - S_M}{\mu_{M+1}},
\end{equation}
where $0 = t_0 < t_1 < \ldots < t_M < t$ are the distances to the different material boundaries in the path of the particle and $\mu_i$ are the mass attenuation coefficients of the corresponding materials:
\begin{equation}
 \mu(\mb x_0 + t\mb \omega) = \mu_i \mbox{ for all } t \in (t_{i-1},t_i).
\end{equation} 
We may therefore generate random variables $T$ with distribution \eqref{distribution-of-free-path-lengths} using an uniform random number generator and Geant4
as described in algorithm \ref{algorithm-monte-carlo}.

\subsection{Geant4 navigation\label{section-geant4-navigation}}

In Geant4 terminology \emph{navigation} is the part of the toolkit that can solve the following simply-stated, interrelated, geometrical problems:
\begin{itemize}
 \item If a particle is in point $A$, inside which geometrical objects, \emph{volumes}, is it?
 \item Will the particle hit a volume boundary, that is, will it exit a volume or enter a new one on its way to point $B$? If so, where, and what is the normal of the boundary surface at that point?
\end{itemize}
These are subproblems for calculating the track of a particle that can
interact with the different materials it passes through. To solve
these problems efficiently to calculate the full track, one stores the
\emph{navigation history} of a particle. The navigation history
describes the volumes, inside which the particle currently is. A
general Geant4 geometry comprises a hierarchy of nested volumes and
the navigation history may indeed contain many levels.  The full
state, including the navigation history, of the particle is stored in
the \emph{navigator}, which also has other information, such as the
normals of the last crossed surface boundaries.

\subsection{Porting Geant4 Navigation to CUDA}

The involved parts of Geant4 were first isolated from the full Geant4
code. They were subsequently translated from C++ to C99 by replacing
C++ classes with C \texttt{structs}, their methods with functions with
\texttt{this}-pointers as parameters and replacing C++ standard
library containers with C-arrays. C++ virtual function were
implemented using ``fork'' functions that call the appropriate
implementation based on the object type. This was because of the lack
of function pointer support in the CUDA version we used.
The reason to translate the program to C was to allow the program to
be also compiled as OpenCL.

\subsection{Benchmark Program \label{section-the-program}}

The basic operation principle of the benchmark program is listed as algorithm \ref{algorithm-main}. The execution phase (\ref{main-pseudo-parallel}) is timed.
All random numbers (in phase \ref{main-random}) are generated using the Mersenne Twister\cite{matsumoto1998mersenne} pseudo-random number generator from the Boost C++ library.
The CPU version of the program can be described as leaving out the transfer
phases \ref{main-pseudo-t0} and \ref{main-pseudo-t1}. The
single-threaded CPU version of the program does phase
\ref{main-pseudo-parallel} sequentially. A multi-threaded CPU version
that uses OpenMP\cite{openmp} to parallelize phase
\ref{main-pseudo-parallel} is also implemented. The GPU version of the
program uses CUDA.

\begin{algorithm}
 \caption{Basic phases of the benchmark program\label{algorithm-main}}
\begin{enumerate}
 \item Load the geometry and transfer it to GPU memory.
 \item Generate $N$ particles: directions and initial positions $(\mb \omega, \mb x_0)_i, i=1,\ldots,N$.
 \item Generate an (exponentially distributed) ``lifetime'' number $Y_i$ for each particle. \label{main-random}
 \item Transfer initial particle states and lifetimes to GPU memory. \label{main-pseudo-t0}
 \item Calculate the free path lengths $T_i$ from $Y_i, (\mb \omega, \mb x_0)_i$ as in algorithm \ref{algorithm-monte-carlo} for each particle in parallel. \label{main-pseudo-parallel}
 \item Transfer the free path length data $T_i$ back to host memory. \label{main-pseudo-t1}
\end{enumerate}
\end{algorithm}

In this program, the attenuation
coefficients $\mu$ are calculated from the densities $\rho$ in the
model geometry as
\begin{equation}
 \mu(\mb x) = \rho(\mb x) \mu_{\mathrm{Al},E},
\end{equation}
where $\mu_{\mathrm{Al},E}$ is the mass attenuation coefficient of
photons in aluminum for given photon energy $E$.
In our tests, all particles have the same energy, 1 MeV, that, according to \cite{xcom}, corresponds to a mass attenuation coefficient $\mu_{\mathrm{Al},E} = 0.06146$ cm$^2$/g.

\subsection{Test inputs}

We use two different test geometries and two different particle distributions
in order to demonstrate how the relative performances of CPU and GPU versions
of the same code can have heavy input dependence.
In both cases we have $N=100000$ input particles.

In the \emph{Random} particle setup the directions of the particles, $\mb \omega$, are uniformly
distributed on the unit sphere $S^2$. The lifetimes of the particles are generated from the exponential distribution.
The output of the program is a random sample from the distribution of the free path lengths of the particles.
In the \emph{Regular} setup, the particles travel towards a regular 400$\times$250 grid and are input to the program in column-major order. Instead of generating Exp(1) distributed lifetimes, all particles
are assigned an equal lifetime of 1. In practice, the output in this case is a 400$\times$250 volume raytraced bitmap image of the geometry.

The first test geometry, \emph{Spheres}, consists of 8000 aluminium balls placed on a regular grid
in a 5$\times$5$\times$5 meter vacuum cube. All particles start from the center of the geometry.
The other, more complex, geometry, \emph{CMS}, is based on a model of the CMS
detector \cite{cms}. It is constructed by extracting all supported
types of solids, namely boxes, tubes and cones, from the original
model.
The densities of the materials in
the CMS model are also imported to the program.

\section{Results}

The GPU version of the program described in Section
\ref{section-the-program} was tested on a Linux desktop with an NVIDIA
GeForce 470 GTX GPU. Both CPU programs, the single- and multi-threaded
versions, were run on a 12-core AMD CPU system (2 $\times$ 6-Core AMD Opteron 2427).
The execution times of the program in different test cases are shown in Table \ref{table-results}.
In all test cases, the GPU was faster than a single CPU core. However, the speedup varied dramatically depending on the input. In the most beneficial example, the GPU was over 13 times faster than the single-threaded CPU version and 60\% faster than the 12-core CPU system. In the most realistic example (CMS, Random), the GPU was only 60\% faster than the single-threaded CPU version and the 12-core system performed 7 times better than this.

\begin{figure}[h]
\begin{center}
 \label{figure-energy-consumption}
  \includegraphics[width=0.8\textwidth]{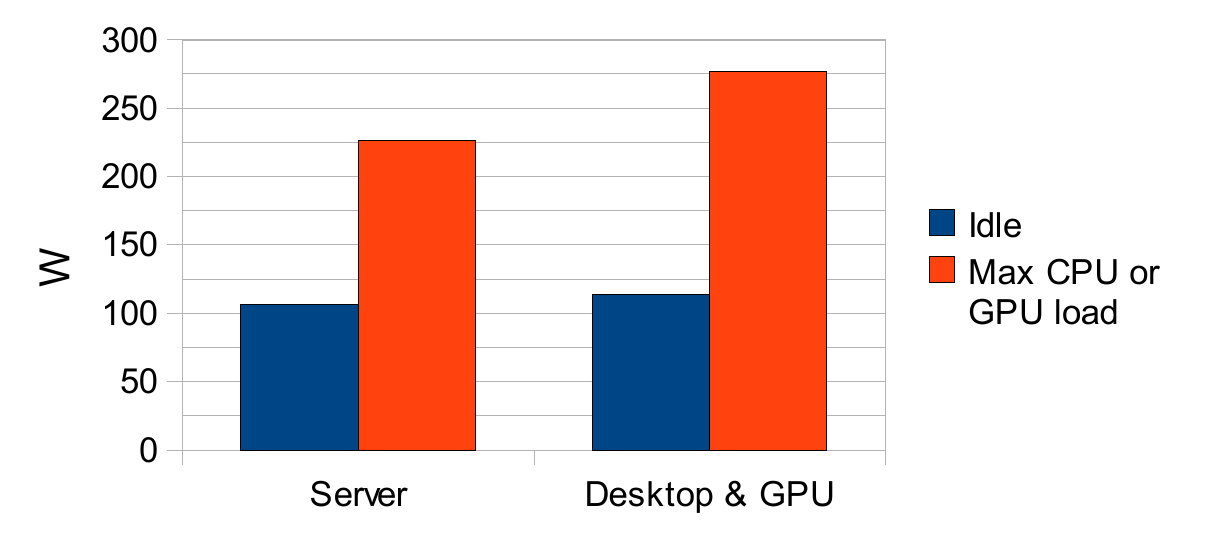}
\end{center}
\caption{Power usage of test hardware}
\label{energy}
\end{figure}

In addition to the run times, we also measured energy usage of the
test hardware. The results are shown in Figure \ref{figure-energy-consumption}. The idle
power consumption of GPU is around 20 W that means around 25\%
increase. The peak power of GPU is around
165W that more than doubles the consumption of the desktop.

\begin{table}
 \caption{Performance of the benchmark simulation\label{table-results}}
\begin{center}
\begin{tabular}{l|l|l|c|c}
Geometry & Particles & Platform & Exec. time (ms) & Speedup factor \\
\hline
 \multirow{3}{*}{Spheres}
 & \multirow{3}{*}{Regular} &
	CPU, 1-thread	& 442 & 1 \\
 &&	CPU, 12-thread	& 54 & 8.2 \\
 &&	GPU		& 33 & 13.3 \\
\hline
 \multirow{3}{*}{Spheres}
 & \multirow{3}{*}{Random} &
	CPU, 1-thread	& 458 & 1 \\
 &&	CPU, 12-thread	& 56 & 8.2 \\
 &&	GPU		& 114 & 4.0 \\
\hline
 \multirow{3}{*}{CMS}
 & \multirow{3}{*}{Regular} &
	CPU, 1-thread	& 22431 & 1 \\
 &&	CPU, 12-thread	& 1862 & 12.0 \\
 &&	GPU		& 8881 & 2.5 \\
\hline
 \multirow{3}{*}{CMS}
 & \multirow{3}{*}{Random} &
	CPU, 1-thread	& 15493 & 1 \\
 &&	CPU, 12-thread	& 1316 & 11.8 \\
 &&	GPU	& 9977 & 1.6 \\
\end{tabular}
\end{center}
\end{table}

\section{Discussion}

Execution path divergence within a warp of threads is a key performance factor on the SIMT architecture~\cite{cuda}.
In theory, heavy divergence can potentially slow down the program by up to a factor 32.
The level of actual run-time divergences (that determines the performance) depends not only on the number of branching instructions in the program, but also on its input.

Geant4 Navigation code is complex and heavily branching and thus as such not optimal for the SIMT architecture. For regular enough input, however, the number level of divergence is low and the program performs well on the GPU. Our more realistic inputs do not exhibit such regularity and the corresponding GPU performance is poor.
The performance is also be affected by other factors than branching, such as register space, memory access patterns and double precision floating point performance. CUDA profiler results showed that register space constraints were indeed a bottleneck factor limiting the performance of our application. This is probably due to the relatively large per-particle Geant4 navigator data structures required in our implementation.

We also tried more complex load balancing schemes on the GPU, such as using CUDPP\cite{cudpp} to periodically remove empty slots (of finished particles) from the warps with parallel stream compaction operations and refilling the GPU with new particles to maximize occupancy. Again, depending on the test case, this resulted to some speedup, but no dramatic performance increases were observed.

GPU implementations of Monte Carlo particle physics simulations have
also been studied by \cite{jahnke2008gpu} and
\cite{jia2010development}. Jahnke et.al. \cite{jahnke2008gpu} have
implemented a Geant4-based software for radiotherapy calculations on
GPUs using CUDA. They report a 10--30 times speedup compared to a
single core CPU when using an NVIDIA GTX 8800 graphics card. Jia et
al. \cite{jia2010development} have implemented a dose planning method
package, also for radiotherapy, in CUDA, and report 5--7 times speedups
when an NVIDIA Tesla C1060 GPU is compared against a 2.27GHz Intel
Xeon CPU.

Our results are not that promising. Different test hardware can
explain results a lot, but it also seems that efficient implementation
of GPU algorithms is also essential. The problem in this approach for hardware
benchmarking is that it is very difficult to compare the performance
of different algorithms on different hardware.

It seems that an efficient GPU implementation of Geant4 in this context would indeed require rewriting the algorithms for the target GPU architecture, instead of porting code to CUDA as such.
Especially, our results show, that it is not sufficient for an algorithm to be easily parallelizable on CPU (MIMD) architectures to yield an efficient GPU implementation.

\section{Conclusions}

We ported a piece of code from a large particle physics simulation
toolkit to a general purpose GPU and compared its performance in a GPU
and a 12 core CPU server.

The results indicated that the performance on the GPU may be heavily input-dependent. In our test cases, the GPU program, when compared to a single-threaded CPU version, was faster by a factor 1.6 -- 13, depending on the test case. The corresponding speedup factor on the 12-core CPU system was 8--12 and it outperformed the GPU in most cases.

The algorithms were originally
designed for CPUs and quite straightforwardly ported to GPU.
Our results indicated that an efficient GPU implementation of Geant4 in the particle physics context would probably require rewriting the algorithms for the target GPU architecture.

Based on our case, we can say that GPU computing can be a
cost-effective solution for simple enough problems but accelerating complex software packages efficiently using GPGPU is not necessarily a feasible alternative, even if the software was easily parallelizable on CPU systems.

\section*{Acknowledgments}

We would like to thank John Apostolakis for proposing Geant4 as a relevant subject of study and providing significant help and ideas for this research.
We gratefully acknowledge the support and cooperation we received for this paper from Dr. Pierre-Yves Burgi and Dr. Benoit Colle of the University of Geneva, and Dr. Paul Albuquerque of HEPIA, Geneva.

\bibliographystyle{plain}
\bibliography{g4gpu}

\end{document}